\documentstyle[seceq,preprint,epsbox]{jpsj}
\input epsf.tex

\newcommand{\eql}{ < \kern -12pt \lower 5pt \hbox{$\displaystyle =$}}
\newcommand{\eqg}{ > \kern -12pt \lower 5pt \hbox{$\displaystyle =$}}
\newcommand{\lsim}{{ < \kern -11.2pt \lower 4.3pt \hbox{$\displaystyle \sim$}}}
\newcommand{\gsim}{{ > \kern -11.2pt \lower 4.3pt \hbox{$\displaystyle \sim$}}}

\def\runtitle{Crossover Temperature from Non-Fermi Liquid to Fermi Liquid 
Behavior in Two Types of Impurity Kondo Model}
\def\runauthor{Satoshi {\sc Yotsuhashi} and Hideaki {\sc Maebashi}}

\title{Crossover Temperature from Non-Fermi Liquid to Fermi Liquid Behavior 
          in Two Types of Impurity Kondo Model}

\author
{Satoshi {\sc Yotsuhashi}\footnote{E-mail: yotsu@eagle.mp.es.osaka-u.ac.jp} 
and Hideaki {\sc Maebashi}$^{1}$}
\inst
{Division of Materials Physics, Department of Physical Science, Graduated School of Engineering Science, Osaka University, Toyonaka, Osaka 560-8531  
\\
$^{1}$Department of Condensed Matter Physics, 
The Institute of Scientific and Industrial Research, 
Osaka University, 8-1 Mihogaoka, Ibaraki, 
Osaka 567-0047
}

\recdate{\today}

\abst
{Numerical renormalization-group results on entropy 
of the anisotropic two-channel Kondo model 
with the band-width cutoff ($D$) 
in the presence of a magnetic field ($h$) 
are obtained to determine crossover temperature  
from the non-Fermi liquid to Fermi liquid fixed point. 
It is found that the crossover temperature 
($T_{\rm x}$) is given by 
$T_{\rm x} \equiv {r} T_{\rm K} 
\sim D(\Delta J/J_{\rm av})^2 e^{-1/J_{\rm av}}$ 
when $(h /T_{\rm K})^2 \ll r \ll 1 $, 
where $T_{\rm K}$, $J_{\rm av}$ and $\Delta J$ are 
the Kondo temperature, 
the average and difference of the exchange coupling constants, respectively. 
This result indicates that non-Fermi liquid behavior can be seen  
even if $\Delta J > T_{\rm K}$.
Robust similarities of the crossover behavior 
in the region around the non-Fermi liquid critical point to 
that of the two-impurity Kondo model are also discussed.}

\kword
{numerical renormalization-group method, two-channel Kondo model, 
two-impurity Kondo model, non-Fermil liquid}

\begin{document}
\sloppy
\maketitle

\section{Introduction}\label{section1}
Non-Fermi-liquid (NFL) behavior observed in some heavy fermion compounds 
and High $T_{\rm c}$ cuprates has stimulated intense studies for a variety of 
impurity models which have NFL quantum-critical 
points.\cite{2Noz80,2Cox87,2Jones88,PVR93,GVRN93}
While each of the models has its own energy scale $T_{\rm K}$ 
below which the thermodynamic and transport properties are geverned by 
the NFL fixed point, 
relevant perturbation which is normaly present in real systems 
introduces another energy scale 
$T_{\rm x}$ where crossover to the Fermi-liquid (FL) fixed point occurs. 
In other words, 
the NFL behavior can be observed 
only when $T_{\rm x}/T_{\rm K} \ll 1$. 
Since real systems would not be exactly on the critical points, 
it is important to know about the above condition that NFL behavior occurs.
In this paper, we investigate the crossover energy scale $T_{\rm x}$ 
so that the condition $T_{\rm x}/T_{\rm K} \ll 1$ is specified 
in a microscopic sense
by using the numerical renormalization-group (NRG) 
method\cite{2Wilson75,2Krish80,2Sakai89} 
for two types of the quantum impurity model, 
the two-channel Kondo model in the presence of a magnetic field and 
the two-impurity Kondo model, with a special emphasis on the effect 
of channel anisotropy.

\begin{figure}[htbp]
\begin{center}
\epsfxsize=8cm \epsfbox{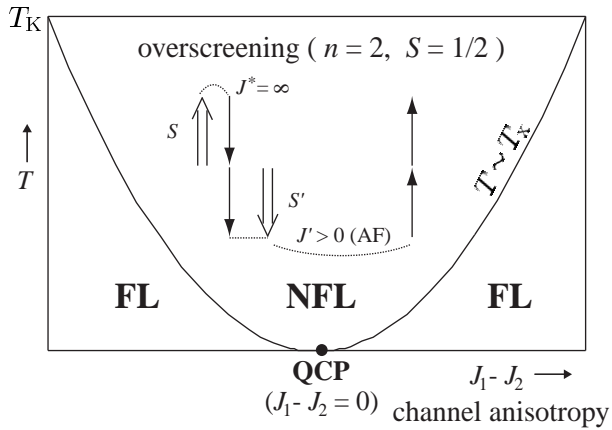}
\end{center}
\caption{A schematic phase diagram near a quantum-critical point (QCP)
of the $n=2$, $S=1/2$ channel anisotropic Kondo model. 
In the region $T_{\rm x} \ll T \ll T_{\rm K}$, 
NFL behavior occurs due to the overscrening Kondo effect.}
\label{fig1}
\end{figure}

The non-Fermi-liquid fixed point of the multichannel Kondo model 
in the overscreened case, i.e., 
when the number of channels $n$ is larger than twice
the size of impurity spin $S$, 
is characterized by a degeneracy of 
the ground state which cannot be completely lifted by coupling 
to conduction electrons.\cite{2Noz80,2Cox} 
The residual entropy is given by 
$\ln [\sin [(2S+1) \pi /(n+2)] / 
\sin [\pi /(n+2)]]$.\cite{AD84,Tsvelick85}
In temperature $T$ below the Kondo temperature $T_{\rm K}$,  
the specific heat coefficient $\gamma$ and magnetic susceptivility $\chi$ 
are $\sim T^{4/(n+2)-1}$, 
the electrical resistivity $\rho$ is 
$const. \pm T^{2/(n+2)}$.\cite{AD84,Tsvelick85,AL91,AL93}
A magnetic field or channel anisotropy is relevant so that it brings about 
crossover to Fermi-liquid fixed points 
but $\gamma$, $\chi$ and $\rho$ keep the above scaling laws when 
$T_{\rm x} \ll T \ll T_{\rm K}$. 
Fig.\ref{fig1} presents a schematic phase diagram 
in the case of $n=2$, $S=1/2$ in which 
$\gamma$, $\chi \sim |\ln T|$ and $\rho \sim const. \pm \sqrt{T}$ 
in the region $T_{\rm x} \ll T \ll T_{\rm K}$.
The two-channel Kondo model has been regarded as one of the most probable 
candidates responsible for the NFL behavior observed in U alloys.\cite{2Cox87} 
Recently, the multichannel Kondo model with channel anisotropy 
has been discussed in relation to singular effects of impurities 
in the region around the ferromagnetic quantum-critical point.\cite{2Mae02}

%..figure 1

%.
%%..T_x in the case of the two-channel Kondo model
%.

The crossover behavior to the FL fixed point 
in the $n=2$, $S=1/2$ Kondo model with channel anisotropy 
and/or in the presence of a magnetic field
has been investigated by a variety of methods.\cite{SS91,2Pan91,2Aff92a,2Fab94,
2Fab95a,2Fab95b,2And95,2Col95}
It can be easily seen in the impurity entropy 
which decreases from $\ln \sqrt{2}$ to $0$ around $T_{\rm x}$ 
as temperature is lowered.
The crossover temperature $T_{\rm x}$ is proportional to the square of 
relevant perturbation as follows: 
\begin{subequations}
\begin{equation}
T_{\rm x} \propto h^2, 
\label{eq:Txa}
\end{equation}
in the presense of a magnetic field $H$ where $h = \mu_{\rm B} H$; 
\begin{equation}
T_{\rm x} \propto (J_1-J_2)^2, 
\label{eq:Txb}
\end{equation}
in the channel anisotropic case 
where $J_1$ and $J_2$ are the exchange couplings for two channels, 
respectively.\cite{2Pan91,2Aff92a}
\end{subequations}
On the other hand, 
the impurity free energy $F_{\rm imp}$ has the following scaling form: 
\begin{equation}
F_{\rm imp}(r;T,h) = - T f(r, T/T_{\rm K}, h/T), 
\label{eq:scaling}
\end{equation}
where $r$ is a dimensionless function of $D$, $J_1$ and $J_2$ 
which depends on a cutoff scheme and in general unknown.\cite{2And95}
Since $r$ represents channel anisotropy, 
$r = 0$ if $J_1=J_2$. 
From eqs. (\ref{eq:Txa}) and (\ref{eq:scaling}), 
we can easily see that $T_{\rm x} \sim h^2/T_{\rm K}$ 
so that crossover from NFL to FL regime occurs if 
$(h/T_{\rm K})^2 \ll 1$ 
in the isotropic channel case.\cite{2Cox,SS91,2Aff92a} 
In the case of channel anisotropy, however, 
eqs. (\ref{eq:Txb}) and (\ref{eq:scaling}) tell us nothing 
about the condition that the NFL behavior can be seen. 
Our main purpose is to identify this condition 
analyzing temperature dependence of the entropy by
the NRG method 
and it will be emphasized that 
crossover from the NFL to FL regime can occur 
even if $|J_1-J_2|>T_{\rm K}$.

%.
%%.. two-impurity systems
%.

The two-impurity Kondo model is also known to have a NFL critical point, 
which arises from competition between the interimpurity 
antiferromagnetic interaction and single-channel Kondo effects.\cite{2Jones88} 
The nature of criticality of this system has been well understood. 
\cite{2Jones88,2Jones89,2Sakai92,2Aff92,2Aff95}
The residual entropy is $\ln \sqrt{2}$ at the critical point just like 
the $n=2$, $S=1/2$ single impurity Kondo model 
which was discussed in the above. 
The staggered impurity susceptibility $\chi_{\rm s}$ is $\sim |\ln T|$ 
when $T_{\rm x} \ll T \ll T_{\rm K}$.
Recently, it has been shown that a similar mechanism works in the model with 
the competition between the Kondo singlet and 
$f^2$-crystalline-electric-field singlet  
explaining the NFL behavior observed 
in R$_{1-x}$U$_x$Ru$_2$Si$_2$ 
(R=Th, Y and La, $x\le 0.07$).\cite{2yotsu01,2yotsu02}

%.
%%..T_x in the case of the two-impurity Kondo model
%.

The crossover temperature $T_{\rm x}$ of the two-impurity Kondo model 
is known to be proportinal to $(K_{\rm c}-K)^2$ 
where $K$ is the interimpurity antiferromagnetic (RKKY) coupling constant 
and $K_{\rm c}$ is the critical coupling.\cite{2Aff95}
We will also elucidate an explicit expression of 
the crossover temperature of this model 
and discuss similarities of the crossover behavior 
to that of the two-channel anisotropic Kondo model 
in a magnetic field.

%. 
%%..Organization
%.
The organization of this paper is as follows. 
In \S \ref{2s}, we discuss effects of channel anisotropy and 
a magnetic field in the two-channel Kondo model.
In \S \ref{3s}, 
we study the crossover behavior of the two-impurity Kondo model 
around the criticality.
Finally, we summarize our results in \S \ref{4s}.

\section{Anisotropic Two-Channel Kondo Model 
in a Magnetic Field}\label{2s}
\subsection{Model and numerical procedure}
%.
%%..Hamiltonian
%.
In this section, 
we discuss crossover from the NFL to FL regime
in the $n=2$, $S=1/2$ Kondo model with channel anisotropy and 
in a magnetic field 
described by the following Hamiltonian:
\begin{eqnarray}
& & H = H_{\rm K}+H_{\rm I}+H_{\rm Z},
\label{eq:model1}\\
& &H_{\rm K}=\sum_{m=1,2}\sum_{k,\sigma} \epsilon_{k}
c_{km\sigma}^\dagger  c_{km\sigma},
\label{eq:2HK1}\\
& &H_{\rm I}=\sum_{m=1,2} \sum_{k,k',\sigma,\sigma'} J_m
c_{km\sigma}^\dagger \vec{\sigma}_{\sigma\sigma'}
c_{k'm\sigma'}\cdot \vec{S}, 
\label{eq:2HI1}\\
& &H_{\rm Z}=-2 h S_{z} .
\end{eqnarray}
Here $m$ ($=1,2$) is the index of two channels, $\sigma$ is the spin index, 
$c_{km\sigma}^\dagger$ is the creation operator of the 
conduction electron, 
$\vec{S}$ is the impurity spin and $\vec{\sigma}$ is the vector 
formed by the Pauli matrices.
$H_{\rm Z}$ represents the Zeeman term 
where the $g$-factor is taken as $2$.
When $J_1$=$J_2$ and $h$=$0$, one obtains the isotropic two-channel Kondo 
Hamiltonian leading to the NFL stable fixed point. 
The case of $J_1$$\neq$$J_2$ 
corresponds to the anisotropic two-channel Kondo Hamiltonian 
whose stable fixed point is that of FL. 

To analyze the Hamiltonian eq. (\ref{eq:model1}) 
by the NRG method,\cite{2Wilson75,2Krish80,2Sakai89} 
we transform eqs. (\ref{eq:2HK1}) and (\ref{eq:2HI1}) into
\begin{eqnarray}
& &H_{\rm K}=\frac{(1+\Lambda^{-1})D}{2}\sum_{m=1,2} \sum_{n=0}^{\infty} 
\sum_{\sigma} \Lambda^{-n/2} 
\nonumber \\
& &\quad\quad\quad \times
(f_{m,n\sigma}^\dagger f_{m,n+1\sigma}+f_{m,n+1\sigma}^\dagger f_{m,n\sigma}) 
\label{HK2}, \\
& &H_{\rm I}=\frac{(1+\Lambda^{-1})D}{2}\sum_{m=1,2} \sum_{\sigma,\sigma'} J_m 
 f_{m,0\sigma}^\dagger \vec{\sigma}_{\sigma\sigma'}
 f_{m,0\sigma'}\cdot \vec{S}, 
\label{HI2}
\end{eqnarray}
where $f_{m,n}$ $(f_{m,n}^\dagger)$ is annihilation (creation) operator of 
the conduction electron in the Wannier representation whose extent 
$k_{\rm F}^{-1} \Lambda^{n/2}$. 
$\Lambda$ is a logarithmic discretization parameter in NRG calculation, 
$D$ is half the bandwidth of the conduction electrons. 
Following an usual procedure of the NRG method, we define $H_N$ as 
\begin{eqnarray}
H_N&=&\Lambda^{(N-1)/2} \bigl[\sum_{m}\sum_{\sigma} \sum_{n=0}^{N-1}\Lambda^{-n/2} 
 (f_{m,n\sigma}^\dagger f_{m,n+1\sigma}+{\rm h.c}) \nonumber\\
&+&J_{m} \sum_{\sigma,\sigma'} 
 f_{m,0\sigma}^\dagger \vec{\sigma}_{\sigma,\sigma'}
 f_{m,0\sigma'}\cdot \vec{S} \bigr] \label{HN1}.
\end{eqnarray}
Then, the Hamiltonian (\ref{HN1}) satisfies the recursion relation
\begin{eqnarray}
H_{N+1}&=&\Lambda^{1/2} H_N+ \sum_{m,\sigma}
(f_{m,N\sigma}^\dagger f_{m,N+1\sigma}+f_{m,N+1\sigma}^\dagger f_{m,N\sigma}) .
\end{eqnarray}
By repeated use of the recursive procedure, 
we can compute the eigen states and energy levels of Hamiltonians 
($H_{N}$). 

Once the energy levels of $H_N$ are obtained, 
the entropy can be calculated by the relation
\begin{eqnarray}
S=\beta _N(\langle H_N \rangle -F_N),
\end{eqnarray}
in each $N$-step. 
Here $\langle H_N \rangle$ and $F_N$ are defined as 
\begin{eqnarray}
\langle H_{\rm N} \rangle &=&\frac{{\rm Tr} H_N e^{-\beta _N H_N}}{{\rm Tr} 
e^{-\beta _N H_N}},\nonumber\\
F_N&=&-\frac{1}{\beta _N}\ln Z_N=-\frac{1}{\beta _N} 
\ln({\rm Tr} e^{-\beta _N H_N}),
\end{eqnarray}
where $\beta _N=\Lambda ^{-(N-1)/2}/T$. 
By setting $T\sim T_N\equiv \Lambda ^{-(N-1)/2}$, 
we can obtain temperature dependence of the entropy 
with a good accuracy in NRG calculation. 

In our calculation, we set $\beta_N^{-1} =0.5$ and $\Lambda=3$ and about 600 
total states are retained at each iteration. 
Hereafter, we take the unit of energy as $(1+\Lambda^{-1}) D / 2$.

\subsection{Crossover temperature $T_{\rm x}$}
%. the way to define T_x
In this subsection, we elucidate an explicit expression of 
the crossover temperature $T_{\rm x}$ of the model, eq. (\ref{eq:model1}), 
from NRG results on the impurity entropy $S_{\rm imp}(T)$. 
As mentioned in \S \ref{section1}, 
if $0$$<$$T_{\rm x}/T_{\rm K}$$\ll$$1$, 
$S_{\rm imp}(T)$ of this model 
decreases from $1/2 \cdot \ln 2$ to $0$ 
around $T_{\rm x}$ as $T$ is lowered.
So $T_{\rm x}$ can be defined 
as temperature at which the entropy takes a value of 
$S_{\rm imp} (T_{\rm x})$=$1/4 \cdot \ln 2$, 
the middle point between NFL and FL fixed points. 
(For examples, see Fig. \ref{fig2} below.)

%..figure 2
\begin{figure}[htbp]
\begin{center}
\epsfxsize=16cm \epsfbox{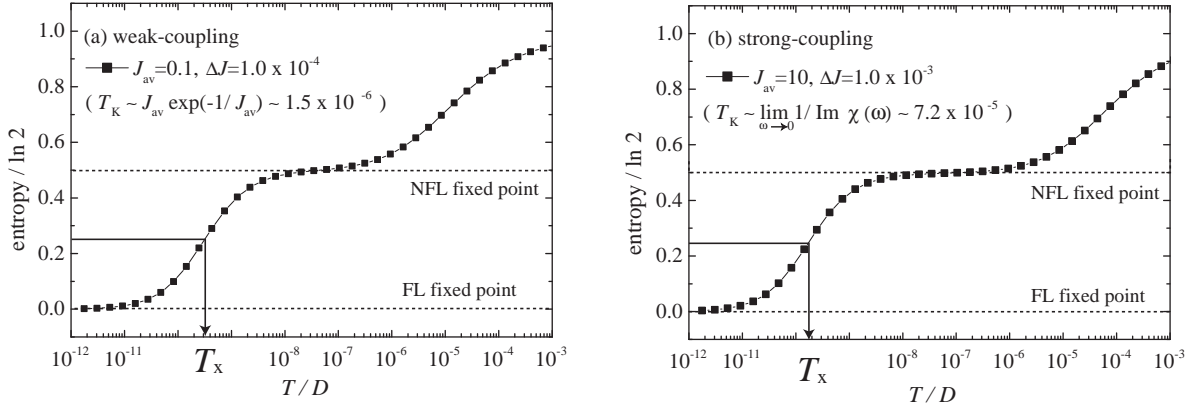}
\end{center}
\caption{NRG results on the entropy $S_{\rm imp}(T)$ 
of the anisotopic two-channel Kondo model for  
(a) $J_{\rm av}=0.1$, $\Delta J = 1.0 \times 10^{-4}$ and  
for (b) $J_{\rm av}=10$, $\Delta J = 1.0 \times 10^{-3}$. 
In each case, crossover from NFL to FL regime can be clearly seen, 
i.e. $T_{\rm x}/T_{\rm K}$$\ll$$1$, 
and then $T_{\rm x}$ is obtained by the relation 
$S_{\rm imp}(T_{\rm x})=1/4 \cdot \ln 2$.}
\label{fig2}
\end{figure}

\begin{figure}[htbp]
\begin{center}
\epsfxsize=16cm \epsfbox{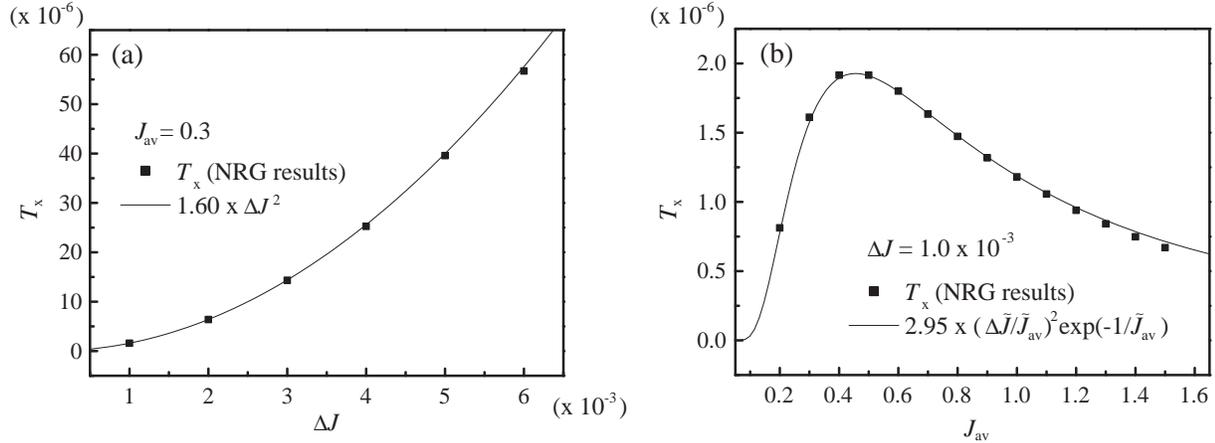}
\end{center}
\caption{Crossover temperature ($T_{\rm x}$) of 
the two-channel Kondo model with channel anisotropy ($\Delta J$).
(a) $T_{\rm x}$ versus $\Delta J$ with $J_{\rm av}$=$0.3$ fixed. 
(b) $T_{\rm x}$ versus $J_{\rm av}$ with $\Delta J$=$1.0\times 10^{-3}$ fixed. 
$T_{\rm x}$ is well fitted by eq.(\ref{2Txc}).}
\label{fig3}
\end{figure}

\begin{figure}[htbp]
\begin{center}
\epsfxsize=16cm \epsfbox{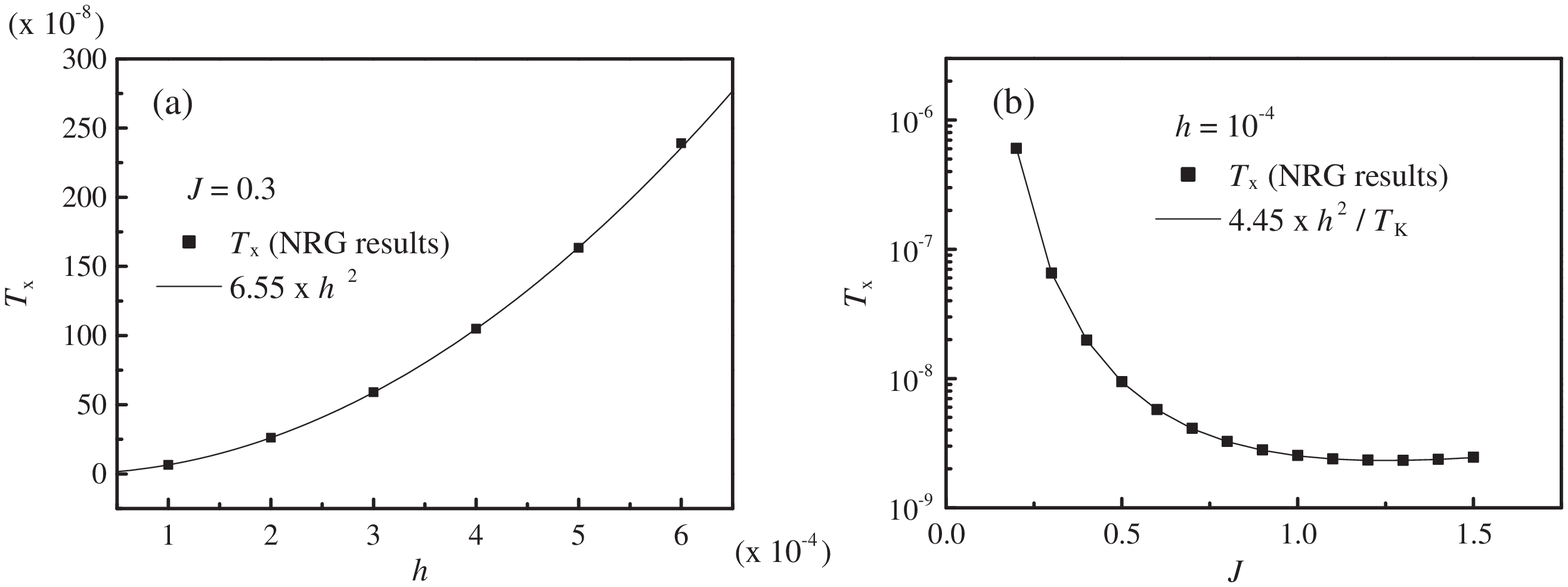}
\end{center}
\caption{Crossover temperature ($T_{\rm x}$) 
of the two-channel Kondo model in the presense of a magnetic field ($h$). 
(a) $T_{\rm x}$ versus $h$ with $J$=$0.3$ fixed. 
(b) $T_{\rm x}$ versus $J$ with $h$=$1.0\times 10^{-4}$ fixed.
$T_{\rm x}$ is well fitted by eq. (\ref{2Txh}).} 
\label{fig4}
\end{figure}

Firstly, we investigate an effect of channel anisotropy on 
$T_{\rm x}$ for $h=0$. 
In this case, we can assume that $T_{\rm x}$ is described by 
a function of two parameters, 
i.e., $J_{\rm av} \equiv (J_1+J_2)/2$ and  $\Delta J \equiv J_1-J_2$. 
Our procedure to determine $T_{\rm x}$ is illustrated 
in Fig. \ref{fig2} (a) and (b) 
which present NRG results on $S_{\rm imp}(T)$ 
for $J_{\rm av}=0.1$, $\Delta J = 1.0 \times 10^{-4}$ and 
for $J_{\rm av}=10$, $\Delta J = 1.0 \times 10^{-3}$, 
respectively. 
It is noted that $T_{\rm x}/T_{\rm K} \ll 1$ while $\Delta J \gsim T_{\rm K}$ 
in both (a) and (b). 
These examples lead us to the fact that 
$T_{\rm x}$ is not scaled as $T_{\rm x} \sim \Delta J^2 / T_{\rm K}$. 
(If $T_{\rm x}$ were $\sim \Delta J^2 / T_{\rm K}$, 
$\Delta J$ should be much less than $T_{\rm K}$ 
for $T_{\rm x}/T_{\rm K} \ll 1$.)
Actually, we have tested a bunch of parameter sets and found a function 
which reproduces the behavior of $T_{\rm x}$ as follows:
\begin{eqnarray}
T_{\rm x}= \alpha \times (\Delta \tilde{J}/\tilde{J}_{\rm av})^2 
e^{-1/\tilde{J}_{\rm av}} \label{2Txc},
\end{eqnarray}
where $\alpha$ is a numerical constant which is taken to be $\alpha=2.95$ 
within the accuracy of our calculation, and 
\begin{eqnarray}
\Delta\tilde{J}&=&A_{\Lambda} \Delta J, \\
\tilde{J}_{\rm av}&=&A_{\Lambda} J_{\rm av}, \\
A_{\Lambda}&=&\frac{1}{2}\frac{1+\Lambda^{-1}}{1-\Lambda^{-1}}\ln \Lambda .
\end{eqnarray}
Here $A_{\Lambda}$ is a revision factor of the density of states 
in NRG calculation.
Fig. \ref{fig3} shows that $T_{\rm x}$ is well fitted by eq. (\ref{2Txc}). 
From Fig. \ref{fig3} (a) in which $J_{\rm av}$=$0.3$ is fixed, 
$T_{\rm x}$ can be seen to be proportional to $\Delta J^2$, which is 
consistent with the results of Pang and Cox.\cite{2Pan91} 
On the other hand, from Fig. \ref{fig3} (b) in which 
$\Delta J$=$1.0\times10^{-3}$ is fixed, 
$J_{\rm av}$-dependence of $T_{\rm x}$ can be 
seen to be equivalent to that of the right hand side of eq. (\ref{2Txc}). 
In the weak-coupling case of $J_{\rm av}$$\lsim$$0.4$, 
$T_{\rm x}$ is an increasing function of $J_{\rm av}$ 
as can be seen in Fig. \ref{fig3} (b)
while $T_{\rm K}$ is well known to be  an increasing function 
$\sim J_{\rm av} e^{-1/J_{\rm av}}$ 
in the case of $n=2$ 
by the two-loop perturbative renormalization group theory.
It is noted that if $T_{\rm x}$ were $\sim \Delta J^2 / T_{\rm K}$, 
$T_{\rm x}$ should be a decreasing function of $J_{\rm av}$ 
in the weak-coupling regime. 
(See also Fig. \ref{fig4} (b) for comparison.)

We have checked that eq. (\ref{2Txc}) can be adopted 
in a wide parameter region of $J_{\rm av}$ 
for the $n=2$, $S=1/2$ Kondo model with channel anisotropy. 
In the weak-coupling regime, 
from eq. (\ref{2Txc}) and $T_{\rm K}\sim J_{\rm av} e^{-1/J_{\rm av}}$, 
the condition that crossover from NFL to FL regime occurs, 
i.e. $T_{\rm x}/T_{\rm K}$$\ll$$1$, is given 
explicitly by $\Delta J^2 / J_{\rm av}^3$$\ll$$1$. 
This result indicates that the NFL behavior in the two-channel Kondo model is 
fairly robust against channel anisotropy. 

%..figure 3

%..figure 4
Next, we investigate an effect of a magnetic field 
in the isotropic channel case of $J_1$=$J_2$$\equiv$$J$. 
In this case, as mentioned in \S \ref{section1}, 
$T_{\rm x}$ is $\sim$$h^2/T_{\rm K}$.\cite{2Cox,SS91,2Aff92a} 
As shown in Fig. \ref{fig4} (a) and (b), 
it can be actually checked from our NRG results on the entropy 
that $T_{\rm x}$ can be well fitted by a function of 
\begin{eqnarray}
T_{\rm x} = \beta \times h^2 / T_{\rm K} \label{2Txh},
\end{eqnarray}
where $\beta$ is a numerical constant of $\beta =4.45$. 
Here $T_{\rm K}$ is defined in terms of 
the dynamical impurity susceptibility $\chi (\omega)$ by  
\begin{eqnarray}
T_{\rm K} \propto \lim _{\omega \to 0} \lim _{T \to 0} 
\left[ {\rm Im} \chi (\omega) \right]^{-1}
\label{2Tk},
\end{eqnarray}
for the isotropic channel model 
in the absence of a magnetic field\cite{2Cox88,2Emery92,2Emery93} 
in which the value of $T_{\rm K}$ is normalized to coincide 
with $J e^{-1/J}$ ($\equiv$$T_{\rm K}^p$) at $J$=$0.1$. 
%..figure 5
\begin{figure}[htbp]
\begin{center}
\epsfxsize=8cm \epsfbox{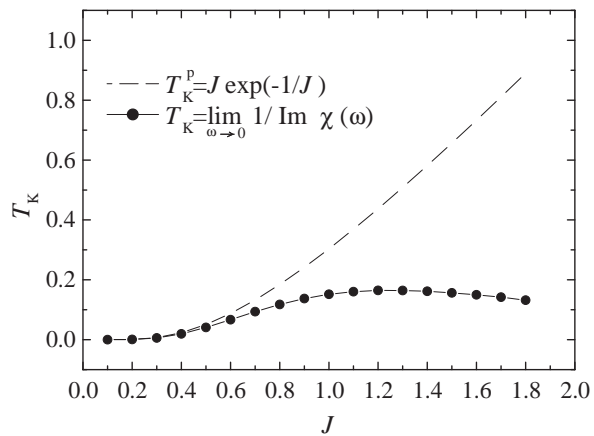}
\end{center}
\caption{$J$-dependence of $T_{\rm K}$ defined by 
eq. (\ref{2Tk}) compared with $T_{\rm K}^p$=$J e^{-1/J}$.}
\label{fig5}
\end{figure}
Eq. (\ref{2Txh}) is valid in a wide region from weak to strong coupling.
This suggests that eq. (\ref{2Tk}) is a good definition of $T_{\rm K}$. 
In Fig. \ref{fig5}, we show $J$-dependence of $T_{\rm K}$, 
which is defined by eq. (\ref{2Tk}), 
and of $T_{\rm K}^p$, 
which is the two-loop perturbative renormalization group result. 
While $T_{\rm K}^p$ is a monotonously increasing function of $J$, 
$T_{\rm K}$ exhibits a broad peak at or around $J$$\sim$$1.3$. 
The discrepancy between $T_{\rm K}$ and $T_{\rm K}^p$ indicates that 
the two-loop perturbative renormalization group result on 
the Kondo temperature is not sufficient for $J\gsim 0.5$. 

\subsection{Scaling function for entropy in the crossover regime}

We have obtained an explicit expression of $T_{\rm x}$, eq. (\ref{2Txc}), 
in the conventional band-width cutoff scheme 
for the $n=2$, $S=1/2$ Kondo model with channel anisotropy. 
On the other hand, 
the thermodynamic Bethe ansatz (TBA) equations for 
the anisotropic multichannel Kondo model was obtained 
by Andrei and Jerez.\cite{2And95} 
While their cutoff scheme differs substantially from ours, 
and so does their expression of $T_{\rm x}$ ($T_{\rm K}$), 
their $n=2$, $S=1/2$ model must be 
in the same universality class as ours. 
It is thus possible to make a comparison of our NRG results with 
the TBA result by use of our expression of $T_{\rm x}$ 
(if necessary, $T_{\rm K}$). 
In this subsection, we concentrate on the crossover regime 
in which $0 \lsim S_{\rm imp}(T) \lsim \ln \sqrt{2}$, $T$$\ll$$T_{\rm K}$, 
and explicitly show that our NRG results on the entropy can be fitted 
by a scaling function of $T$$/$$T_{\rm x}$. 

First, we derive the scaling function in the crossover regime 
analytically from the TBA equations. 
In the scaling limit, 
the impurity part of the free energy is given by\cite{2And95} 

\begin{equation}
F_{\rm imp}(T) = 
-\frac{T}{2 \pi} \int^{\infty}_{-\infty}
\frac{\ln [1+\eta_1(\xi)]}{\cosh [\xi+\ln (T/T_{\rm K})]} 
{\rm d} \xi, \label{2Fimp1}
\end{equation}
where $\eta_1 (\xi)$ is determined by the TBA equations; 
\begin{eqnarray}
& & \ln \eta_1(\xi) = -2 r {\rm e}^\xi 
+ G \ast \ln [1+\eta_2 (\xi)],
\label{2eta1}
\\
& & \ln \eta_2(\xi) = -{\rm e}^\xi + G \ast \ln [1+\eta_1 (\xi)] 
+G \ast \ln [1+\eta_3 (\xi)],
\label{2eta2}
\\
& & \ln \eta_j(\xi) = G \ast \ln [1+\eta_{j-1} (\xi)] 
+ G \ast \ln [1+\eta_{j+1} (\xi)],
\hspace{0.3cm} j \geq 3, 
\label{2eta3}
\end{eqnarray}

\noindent
with a boundary condition 
\begin{equation}
\lim_{j \to \infty} \frac{\ln \eta_j(\xi)}{j} 
= \frac{2 h}{T}.
\end{equation}

\noindent
Here $G \ast \varphi (\xi) \equiv 
\int^{\infty}_{-\infty} G(\xi - \xi')\varphi (\xi') {\rm d}\xi'$ 
and $G(\xi - \xi') \equiv 1/2\pi \cosh (\xi - \xi')$. 
For $T \ll T_{\rm K}$, the impurity free energy, eq. (\ref{2Fimp1}), 
is determined by the $\xi$$\to$$+\infty$ asymptotic form of $\eta_1$. 
In the limit of $\xi$$\to$$+\infty$, $\eta_j$ is given by
\begin{equation}
\eta_j(+\infty) = 
\frac{\sinh^2 [(j-1)h/T]}{\sinh^2(h/T)}-1,
\hspace{0.3cm} j \geq 2. 
\label{eq:limit}
\end{equation}

Firstly, we consider the case of $T$$\ll$$h$$\ll$$T_{\rm K}$. 
By eq. (\ref{eq:limit}), $\eta_3(+\infty) \simeq \exp (2h/T)$ and 
its corrections due to $\xi$ dependence can be neglected, so that 
eq. (\ref{2eta2}) leads to 

\begin{equation}
\ln \eta_2(\xi) = -{\rm e}^\xi + h/T + f(r,{\rm e}^{-\xi},h/T), 
\label{eq:eta2}
\end{equation}

\noindent
where $f$ is defined by eq. (\ref{eq:scaling}). 
In eq. (\ref{eq:eta2}), 
the last term can be seen to compare negligibly with $h/T$
if we note that this term is 
$\sim$$F_{\rm imp} (T)/T$ in a region around $\xi$$\sim$$\ln (T_{\rm K}/T)$. 
Then, after some manipulations 
of eqs. (\ref{2Fimp1}), (\ref{2eta1}) and (\ref{eq:eta2}), 
we obtain 

\begin{equation}
F_{\rm imp}(T) \sim 
-\frac{T}{2 \pi} \int^{\infty}_{-\infty}
\frac{
\ln [1+\exp (- (h/T) \varepsilon (\zeta)) ]}{
\cosh [\zeta+\ln (h/T_{\rm K})]} 
{\rm d} \zeta, 
\label{eq:Fimp1}
\end{equation}

\noindent
where $\varepsilon (\zeta)$ is given by

\begin{eqnarray}
\varepsilon (\zeta) &= & {\rm e}^{\zeta} \left( 
2 r + \frac{1}{2 \pi} \ln \left( 1+ {\rm e}^{-2\zeta} \right)
- \frac{1}{\pi} {\rm e}^{-\zeta} \tan^{-1}\left( {\rm e}^{-\zeta} \right) 
\right),
\\
&\sim & {\rm e}^{\zeta} \left( 
2 r - {\rm e}^{-2\zeta} / 2 \pi \right), 
\hspace{1cm} {\rm for} \quad {\rm e}^{-\zeta} \sim h/T_{\rm K} \ll 1.
\label{eq:vareps}
\end{eqnarray}

\noindent
The integral in eq. (\ref{eq:Fimp1}) has dominant contributions from 
a region around $\zeta \sim \ln (T_{\rm K}/h)$. 
When $r \gg (h/T_{\rm K})^2$, the second term can be neglected 
in eq. (\ref{eq:vareps}) so that the impurity free energy can be 
written in the following scaling form: 
\begin{equation}
F_{\rm imp}(T) = - T {\tilde f} (T/T_{\rm x}), 
\end{equation}
where the crossove temperature is given by $T_{\rm x}=rT_{\rm K}$ and 
${\tilde f}$ is a scaling function in the crossover regime; 
\begin{eqnarray}
{\tilde f} (x)&= &
\int^{\infty}_{-\infty}
\frac{\ln [1+\exp (-2 \exp(\xi))]}{2 \pi \cosh [\xi+\ln (x)]} 
{\rm d} \xi
\\
&= &\frac{1}{\pi x} \ln \frac{1}{\pi x} - \frac{1}{\pi x} 
- \ln \Gamma \left( \frac{1}{\pi x} + \frac{1}{2}\right) 
+ \frac{1}{2} \ln 2 \pi ,
\end{eqnarray}
where $\Gamma$ is the gamma function. 
Then, the impurity entropy, 
$S_{\rm imp}$=$-\partial F_{\rm imp}/\partial T$, is calculated as 
\begin{eqnarray}
S_{\rm imp}(T) &= & 
\frac{1}{\pi (T/T_{\rm x})} \psi \left( \frac{1}{\pi (T/T_{\rm x})} 
+ \frac{1}{2} \right)
- \frac{1}{\pi (T/T_{\rm x})} 
\nonumber
\\
& & - \ln \Gamma \left( \frac{1}{\pi (T/T_{\rm x})} + \frac{1}{2} 
\right) 
+ \frac{1}{2} \ln 2 \pi, 
\label{2entBA}
\end{eqnarray}
where $\psi$ is the digamma function. 
On the other hand, 
when $r \ll (h/T_{\rm K})^2$, the first term can be neglected 
in eq.(\ref{eq:vareps}). 
In this case, if the crossover temperature is regarded as 
$T_{\rm x}$=$h^2/4 \pi T_{\rm K}$, 
$T$-dependent contributions in the impurity free energy is 
given by $-T{\tilde f}(T/T_{\rm x})$ and therefore 
eq. (\ref{2entBA}) is also available for the entropy.

Secondly, we consider the case of $h$$\lsim$$T$$\ll$$T_{\rm K}$. 
By eqs. (\ref{2eta2}) and (\ref{eq:limit}), 
$\ln \eta_2(\xi)$$\simeq$$- {\rm e}^{-\xi}$ 
so that $\ln \eta_1(\xi)$$\simeq$$- 2 r {\rm e}^{\xi}$. 
Hence eq. (\ref{2entBA}) is available by $T_{\rm x}$=$rT_{\rm K}$. 
It is noted that if $r \ll (h/T_{\rm K})^2$, 
$T_{\rm x}$=$rT_{\rm K}$$\ll$$h$$\lsim$$T$ so that 
the entropy is $\sim$$\ln \sqrt{2}$, which is not sensitive to 
the value of $T_{\rm x}$. 

In summary, a scaling function for entropy in the crossover regime 
is given by eq. (\ref{2entBA}) where 
$T_{\rm x}$=$rT_{\rm K}$ for $r$$\gg$$(h/T_{\rm K})^2$ and 
$T_{\rm x}$=$h^2/4 \pi T_{\rm K}$ for $r$$\ll$$(h/T_{\rm K})^2$.
For eq. (\ref{2entBA}), 
$S_{\rm imp}$$(T_{\rm x})$=$0.99$$\times$$1/4$$\cdot$$\ln 2$ 
so that $T_{\rm x}$ is about the same as we defined in the former subsection. 
(We must apologize for the normalization schemes on $T_{\rm K}$ 
which are different between the former and this subsections.)

\begin{figure}[htbp]
\begin{center}
\epsfxsize=8cm \epsfbox{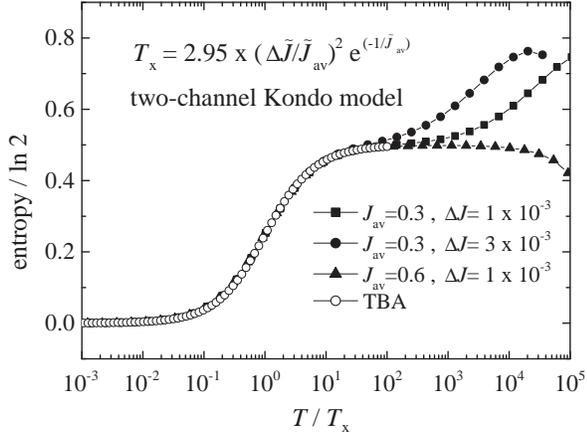}
\end{center}
\vspace{-5mm}
\caption{Temperature dependences of entropy 
obtained by the NRG method for the anisotropic two-channel Kondo model.
In the crossover regime, by use of an expression of $T_{\rm x}$, 
eq. (\ref{2Txc}), they can be fitted quite well 
by a scaling function, eq. (\ref{2entBA}) derived from the TBA equations.}
\label{fig6}
\end{figure}

\begin{figure}[htbp]
\begin{center}
\epsfxsize=8cm \epsfbox{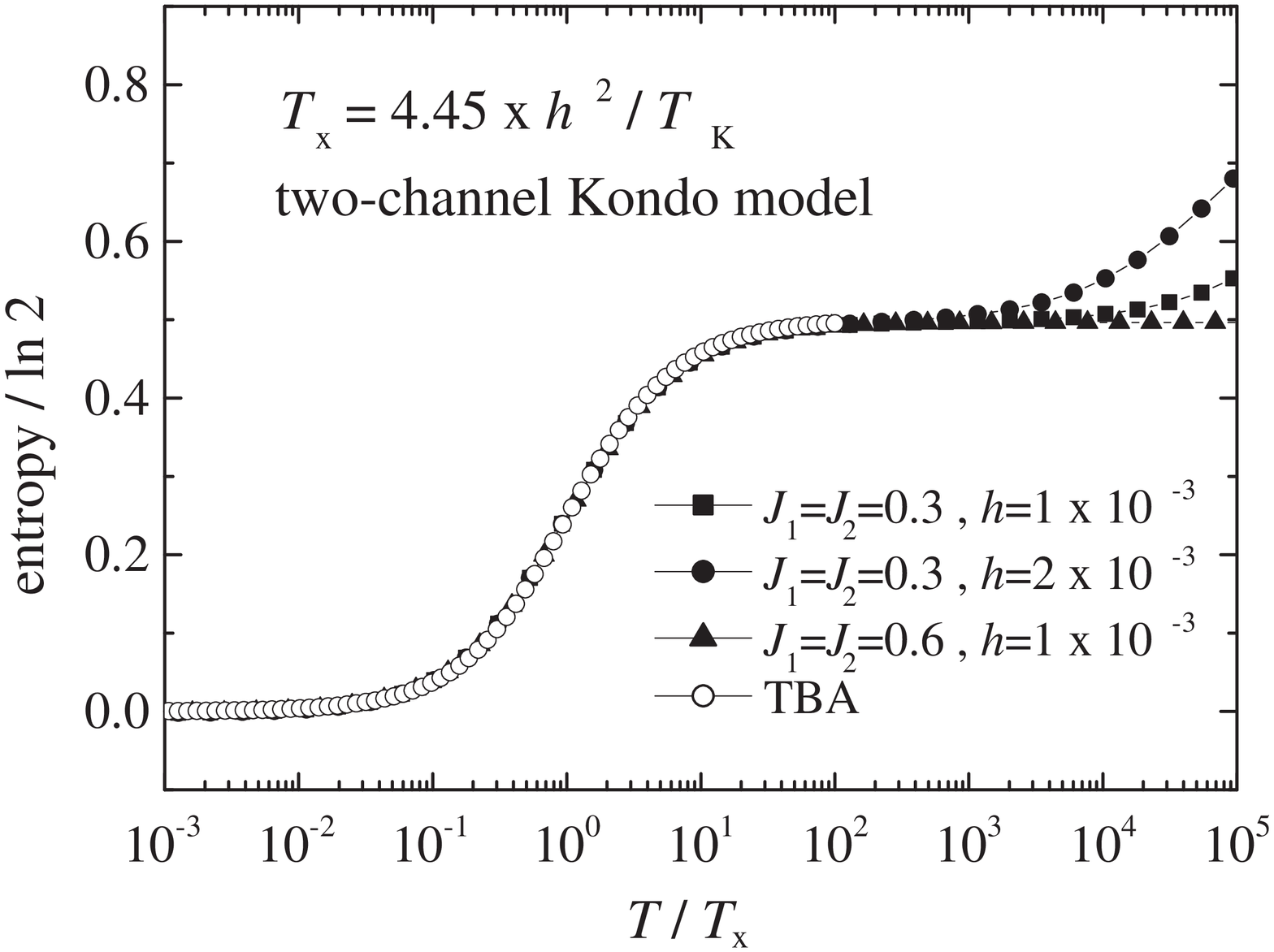}
\end{center}
\vspace{-5mm}
\caption{Temperature dependence of entropy 
obtained by the NRG method for the two-cannel Kondo model 
in the presence of a magnetic field. 
In the crossover regime, by use of an expression of $T_{\rm x}$, 
eq. (\ref{2Txh}), they can be fitted quite well 
by a scaling function, eq. (\ref{2entBA}) derived from the TBA equations.
}
\label{fig7}
\end{figure}

Second, we compare our NRG results on entropy with eq. (\ref{2entBA}).
Figures \ref{fig6} and \ref{fig7} present the results on 
entropy versus temperature scaled by $T_{\rm x}$, 
expressions of which are eq. (\ref{2Txc}) and eq. (\ref{2Txh}),
in the anisotropic channel case and 
in the presence of a magnetic field, respectively.
In both cases, the $T$ dependence of entropy can be fitted quite well 
in the crossover regime by the scaling function, eq. (\ref{2entBA}), 
derived from the TBA equations.

%..figure 6

%..figure 7

\section{Two-Impurity Kondo Model}\label{3s}
\subsection{Model}
%..Hamiltonian
In this section, we study the two-impurity Kondo model whose Hamiltonian 
is given by two independent Kondo models coupled by interimpurity exchange 
interaction ($K$) as follows:
\begin{eqnarray}
&&H=H_{\rm K}+H'_{\rm I}+H_{\rm L} \label{2H1}, 
\\
&&H_{\rm K}=\sum_{m=1,2}\sum_{k,\sigma} \epsilon_{k}
c_{km\sigma}^\dagger  c_{km\sigma} , \label{2HK1}\\
&&H'_{\rm I}=\sum_{m=1,2} J_m\sum_{k,k',\sigma,\sigma'}
c_{km\sigma}^\dagger \vec{\sigma}_{\sigma\sigma'}
c_{k'm\sigma'}\cdot \vec{S}_m , \label{2HI1}\\
&&H_{\rm L}=K \vec{S}_1 \cdot \vec{S}_2 .\label{2HCEF1}
\end{eqnarray}
The notations are the same as before.
In the original form of the two-impurity Kondo model studied by Jones 
{\it et al}.,
\cite{2Jones88} the Kondo exchange coupling between the conduction electrons 
and localized spin with both the same and different symmetry 
(specified by $m$) are taken into 
account, because they considered the case of 
the two impurities on different sites.
In the present study, we neglect 
the Kondo exchange coupling between different symmetries, 
for simplicity. 
It is however noted that 
our model, eq. (\ref{2H1}), itself can be considered to be 
a local-moment version of the two-orbital Anderson model 
with crystalline-electric-field (CEF) effect 
in a certain pseudospin representation.\cite{2yotsu01,2yotsu02} 

The model, eq. (\ref{2H1}) has a NFL fixed point at $K$=$K_{\rm c}(J_1,J_2)$ 
for a set of $J_1$ and $J_2$ due to competition between 
the single channel Kondo effects and the interimpurity antiferromagnetic 
interaction.\cite{2Jones88,2Jones89,2Sakai92,2Aff92,2Aff95,
2yotsu01,2yotsu02,2Kura92}
In this section, we concentrate on the region around critical points of 
$K$=$K_{\rm c}(J,J)$. 
The definition of $T_{\rm x}$ is the same as that of the former section; 
$S_{\rm imp} (T_{\rm x})$=$1/4 \cdot \ln 2$.

\subsection{Crossover temperature $T_{\rm x}$ 
in the region around critical points of $K$=$K_{\rm c}(J,J)$}

\begin{figure}[htbp]
\begin{center}
\epsfxsize=8cm \epsfbox{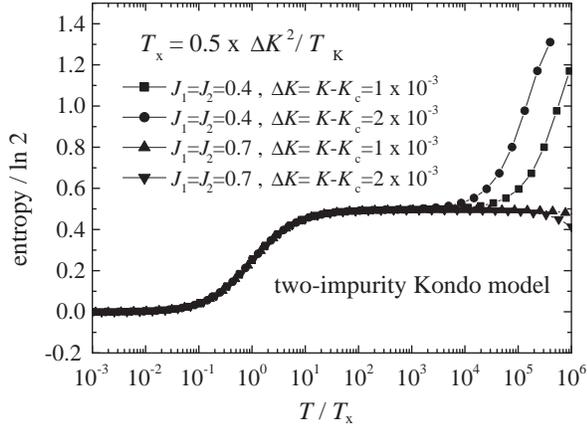}
\end{center}
\vspace{-5mm}
\caption{Temperature dependence of 
entropy for the two-impurity Kondo model 
when $K$ is displaced slightly away from $K_{\rm c}$. 
$K_{\rm c}$=$0.564723$ for $J_1$=$J_2$=$0.4$.
$K_{\rm c}$=$2.655538$ for $J_1$=$J_2$=$0.7$.
The temperature is scaled by $T_{\rm x}$, which is determined by eqs. 
(\ref{2Tkiaf}) and (\ref{2Tx2imp}).
}
\label{fig8}
\end{figure}

Firstly, we consider cases in which $K$ is displaced slightly away from 
a critical value of $K_{\rm c}$$\equiv$$K_{\rm c}(J,J)$ 
with $J_1$=$J_2$=$J$ fixed. 
It has already been discussed by using the conformal field theory 
that $T_{\rm x}$ is proportional to $\Delta K^2 /T_{\rm K}$, \cite{2Aff95}
where $\Delta K$=$K_{\rm c}-K$. 
Here we define $T_{\rm K}$ by
\begin{eqnarray}
T_{\rm K} \equiv 
\lim _{\omega \to 0}\lim _{T \to 0} 1/ {\rm Im} \chi_{\rm s} (\omega) 
\label{2Tkiaf},
\end{eqnarray}
where $\chi_{\rm s}$ is the dynamical impurity  
susceptibility for $\vec{S_1}-\vec{S_2}$. 
We have examined various parameter sets and confirmed that the 
crossover temperature $T_{\rm x}$ is given as
\begin{eqnarray}
T_{\rm x}=\gamma \times \Delta K^2 / T_{\rm K}, \label{2Tx2imp}
\end{eqnarray}
where $\gamma$ is a numerical constant of $\gamma$=$0.50$ 
within the accuracy of our calculation.
In fact, as shown in Fig. \ref{fig8}, the temperature dependence of 
entropy can be scaled by $T_{\rm x}$ defined by eq.(\ref{2Tx2imp}) quite well. 
%..figure 8

\begin{figure}[htbp]
\begin{center}
\epsfxsize=8cm \epsfbox{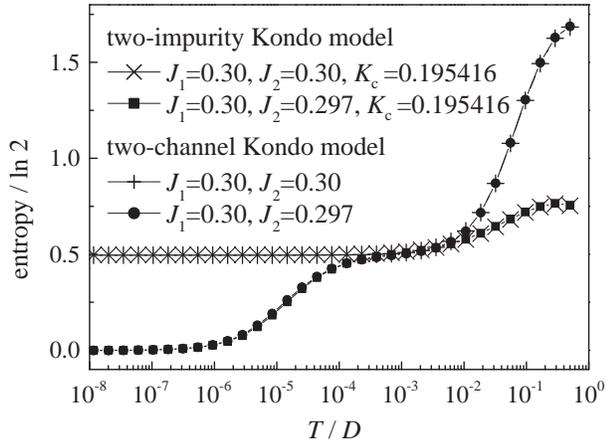}
\end{center}
\vspace{-5mm}
\caption{Crossover behavior for the two-impurity Kondo model 
($J_1$=$0.3$, $J_2$=$0.297$, and $K$=$K_{\rm c}$=$0.195416$), 
compared with the case of the anisotropic two-channel Kondo model with 
$J_1$=$0.3$ and $J_2$=$0.297$.}
\label{fig9}
\end{figure}

Next, we consider cases in which 
$J_2$ is displaced slightly away from $J_2$=$J$ with 
$J_1$=$J$ and $K$=$K_{\rm c}$$\equiv$$K_{\rm c}(J, J)$ fixed.
In Fig. \ref{fig9}, we show the behavior of a crossover for 
$J_1$=$0.3$, $J_2$=$0.297$, and $K$=$K_{\rm c}$=$0.195416$, compared with 
the case of the two-channel Kondo model with $J_1$=$0.3$ and $J_2$=$0.297$.
The isotropic channel cases in both models are also shown in this figure.
In both cases, the small perturbation ($J_1-J_2$) does not affect the 
high temperature behavior.
Namely, it does not change $T_{\rm K}$. 
The temperature dependence of entropy in the anisotropic channel case 
is quite similar in both cases in the crossover regime. 
Hence $T_{\rm x}$ for the two-impurity model is also given by eq. (\ref{2Txc}).

%..figure 9

\section{Summary}\label{4s}

In this paper, we have investigated effects of relevant perturbations, 
which bring about crossover from the NFL to FL regime, 
in the regions around 
quantum critical points of the two-channel and two-impurity Kondo models. 

For the two-channel Kondo model in which $T_{\rm K}$ is defined by 
$T_{\rm K} \propto \lim_{\omega \to 0}\lim_{T \to 0} 
[{\rm Im} \chi (\omega)]^{-1}$ 
(in the weak-coupling regime, 
$T_{\rm K}$$\simeq$$J_{\rm av}{\rm e}^{-1/J_{\rm av}}$), 
we have derived numerically exact expressions of 
the crossover temperature $T_{\rm x}$ 
from our NRG results on the entropy as follows:
(1) in the anisotropic channel case, 
if $T_{\rm x}$/$T_{\rm K}$$\ll$1 
(in the weak-coupling regime, 
$\Delta J^2/J_{\rm av}^3$$\ll$1 ), 
crossover from the NFL to FL regime occurs around $T_{\rm x}$ 
which is given by 
$T_{\rm x}$=$ \alpha $$\times$$ (\Delta \tilde{J}/\tilde{J}_{\rm av})^2 $$
e^{-1/\tilde{J}_{\rm av}}$ with $\alpha$=$2.95$, 
where $J_{\rm av}$ and $\Delta J$ 
are the average and difference of the exchange couplings $J_1$ and $J_2$, 
respectively (a swung dash means $A_{\Lambda}$ times, 
$A_{\Lambda}$ being a revision factor of the order of 1); 
(2) in the presence of a magnetic field, 
if $(h/T_{\rm K})^2$$\ll$1, 
crossover from the NFL to FL regime occurs around $T_{\rm x}$ 
which is given by 
$T_{\rm x}$=$\beta \times h^2 / T_{\rm K}$, 
where $\beta$=$4.45$ in our normalization procedure on $T_{\rm K}$.

For the two-impurity Kondo model which has a NFL critical point 
at $K$=$K_{\rm c}(J_1,J_2)$ 
for a set of the Kondo exchange couplings $J_1$ and $J_2$, 
we have obtained NRG results on the entropy 
in the region around critical points 
of $K$=$K_{\rm c}(J,J)$ as follows: 
(1) when $J_2$ is displaced slightly away from $J_2$=$J$ with 
$J_1$=$J$ and $K$=$K_{\rm c}(J,J)$ fixed, 
crossover to a FL fixed point occurs around 
$T_{\rm x}$ which is given by that of the two-channel Kondo model 
with the same set of $J_1$ and $J_2$ in the absence of a magnetic field;
(2) when interimpurity (RKKY) exchange coupling $K$ is displaced slightly 
away from the critical value of $K_{\rm c}(J,J)$ with $J_1$=$J_2$=$J$ fixed, 
crossover to a FL fixed point occurs around $T_{\rm x}$ which is given by 
$T_{\rm x}$=$\gamma \times (K_{\rm c}-K)^2 / T_{\rm K}$ with $\gamma$=$0.50$ 
where 
$T_{\rm K}$$\equiv$$\lim_{\omega \to 0}$$\lim_{T \to 0}$
${\rm Im} \chi_{\rm s} (\omega)$.

In the present work, 
all the results on $T_{\rm x}$ can be adopted in a wide region of 
$J$ ($J_{\rm av}$) and  
all the temperature dependences of entropy in the crossover regime 
can be fitted quite well by a scaling function, eq. (\ref{2entBA}), 
derived from the TBA equations of the two-channel Kondo model 
with channel anisotropy and in the presence of a magnetic field. 
It is emphasized that the NFL behavior in the two channel Kondo model 
is fairly robust against channel anisotropy so that 
crossover from the NFL to FL regime can be seen 
even if $\Delta J$$>$$T_{\rm K}$.

\section*{Acknowledgments}
We wish to thank C. M. Varma for valuable advice. 
We also would like to thank I. Affleck, D. L. Cox, H. Kohno, H. Kusunose 
and K. Miyake for helpful comments. 
H. M. is supported by Reserch Fellowship of Japan Society 
for the Promotion of Science for Young Scientists. 
This work is supported in part by the Grant-in-Aid for COE Research (10CE2004) 
of the Ministry of Education, Science, Sports and Culture.

\end{document}